\documentclass[journal,twoside,web]{ieeecolor}
\usepackage{tmi}
\usepackage{cite}
\usepackage{amsmath,amssymb,amsfonts}
\usepackage{algorithmic}
\usepackage{graphicx}
\usepackage{textcomp}
\usepackage{amsmath}
\usepackage{wrapfig}
\usepackage{float}
\usepackage{adjustbox}
\usepackage{booktabs}
\usepackage{bm}
\usepackage{wrapfig}
\usepackage[scientific-notation=true]{siunitx}
\usepackage{hyperref}
\definecolor{mymarked}{rgb}{0.0, 0.0, 0.0}
\newcommand{\marked}[1]{\textcolor[RGB]{0, 0, 0}{#1}}
\newcommand{\itm}[1]{{\texttt{\small #1}}}
\newcommand{\ourmethodN}{MedSyn}
\newcommand{\ourmethod}{MedSyn~}
\def\BibTeX{{\rm B\kern-.05em{\sc i\kern-.025em b}\kern-.08em
    T\kern-.1667em\lower.7ex\hbox{E}\kern-.125emX}}
\begin{document}
\title{MedSyn: Text-guided Anatomy-aware Synthesis of High-Fidelity 3D CT Images}
\author{Yanwu Xu, Li Sun, Wei Peng, Shuyue Jia, Katelyn Morrison, Adam Perer, Afrooz Zandifar, Shyam Visweswaran, 
Motahhare Eslami, and Kayhan Batmanghelich
\thanks{This work is equally contributed by Y. Xu, L. Sun and W. Peng and was partially supported by NIH Award Number 1R01HL141813-01, NSF 1839332 Tripod+X, and SAP SE. 
We were also grateful for the computational resources provided by Pittsburgh SuperComputing grant number TG-ASC170024.}
\thanks{Y. Xu, L. Sun, S. Jia and  K. Batmanghelich are with Department of Electrical and Computer Engineering,
Boston University, Boston, MA 02215 (email:  \{yanwuxu,lisun,brucejia,batman\}@bu.edu)}
\thanks{W. Peng are with Department of Psychiatry and Behavioral Sciences,
Stanford University, Stanford, CA 94305 (email:  wepeng@stanford.edu)}
\thanks{K. Morrison, A. Perer and M. Eslami are with Human-Computer Interaction Institute,
Carnegie Mellon University, Pittsburgh, PA 15213 (email:  kcmorris@cs.cmu.edu, adamperer@cmu.edu, meslami@andrew.cmu.edu)}
\thanks{ A. Zandifar is with the  University of Pittsburgh Medical Center, Pittsburgh, PA 15213 (email: zandifara@upmc.edu)}
\thanks{ S. Visweswaran is with the Department of Biomedical Informatics, University of Pittsburgh, Pittsburgh, PA 15206 (email: shv3@pitt.edu
)}
}

\maketitle

\begin{abstract}
This paper introduces an innovative methodology for producing high-quality 3D lung CT images guided by textual information. While diffusion-based generative models are increasingly used in medical imaging, current state-of-the-art approaches are limited to low-resolution outputs and underutilize radiology reports' abundant information. The radiology reports can enhance the generation process by providing additional guidance and offering fine-grained control over the synthesis of images. Nevertheless, expanding text-guided generation to high-resolution 3D images poses significant memory and anatomical detail-preserving challenges. Addressing the memory issue, we introduce a hierarchical scheme that uses a modified UNet architecture. We start by synthesizing low-resolution images conditioned on the text, serving as a foundation for subsequent generators for complete volumetric data. To ensure the anatomical plausibility of the generated samples, we provide further guidance by generating vascular, airway, and lobular segmentation masks in conjunction with the CT images. The model demonstrates the capability to use textual input and segmentation tasks to generate synthesized images. \marked{Algorithmic comparative assessments and blind evaluations conducted by 10 board-certified radiologists indicate that our approach exhibits superior performance compared to the most advanced models based on GAN and diffusion techniques, especially in accurately retaining crucial anatomical features such as fissure lines and airways.} This innovation introduces novel possibilities. This study focuses on two main objectives: (1) the development of a method for creating images based on textual prompts and anatomical components, and (2) the capability to generate new images conditioning on anatomical elements. The advancements in image generation can be applied to enhance numerous downstream tasks.
\end{abstract}


\begin{figure*}[htp]
  \centering
    \includegraphics[width=18cm]{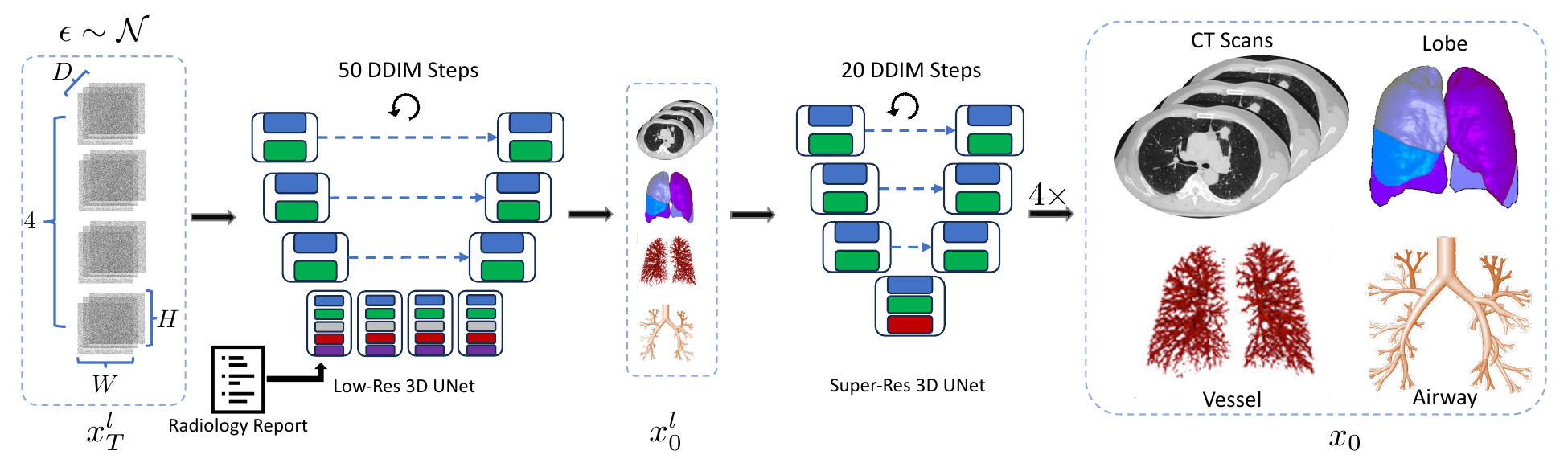}
  \caption{Overview of our generative model, \ourmethodN. Using a hierarchical approach, we first generate a 64$\times$ 64 $\times$ 64 low-resolution volume, along with its anatomical components, conditioning on Gaussian noise $\epsilon$ and radiology report. The low-resolution volumes are then seamlessly upscaled to a detailed 256$\times$ 256 $\times$ 256 resolution.}
  \label{fig:mainSchematic}
\end{figure*}
\section{Introduction}
\label{sec:intro}
Denoising Diffusion Probabilistic Models (DDPM)~\cite{ddpm}, also known as score-based generative models~\cite{score_sde}, have emerged as powerful tools in both computer vision and medical imaging due to their stability during training and exceptional generation quality. State-of-the-art image generation tools, such as Imagen~\cite{IMAGEN} and latent Diffusion Models (LDMs)~\cite{ldm}, can employ text prompts to provide fine-grained guidance during image generation. This capability is promising for medical imaging, encompassing applications like privacy-preserving data generation, image augmentation, black box uncertainty quantification, and explanations. Although methods such as RoentGen~\cite{roentgen} have illustrated the potential of 2D cross-modality generative models conditioned on text prompts, there are no known text-guided volumetric image generation techniques for medical imaging. Extending such an approach to 3D presents challenges, including high memory demands and preserving crucial anatomical details. This paper aims to address these challenges.

To enhance the image resolution for generative models, increasing the voxel count within a fixed field of view is memory-intensive~\cite{hagan2022}.
Synthesizing 3D volumetric images at high resolutions (e.g., $256\times 256 \times 256$) demands substantial memory since the neural network must store intermediate activations for backpropagation. To preserve essential anatomical details and condition on input text prompts, a high-capacity network is essential, further compounding the issue. Although GANs~\cite{shan20183,jin2019applying,cirillo2020vox2vox,hagan2022} and conditional Denoising Diffusion Probabilistic Models (cDPMs)~\cite{sequential_diffusion, sequential_diffusion2, pinaya2023generative} have set benchmarks in volumetric medical imaging synthesis, each has its limitations. GANs, despite their inference efficiency, can compromise sample diversity and sometimes produce anatomically implausible artifacts. Conversely, diffusion models, built to boost sample diversity through iterative denoising, grapple with memory constraints, primarily due to their resource-intensive 3D attention UNet. The iterative denoising coupled with the sequential sub-volume generation in cDPMs makes them time-intensive at the inference stage. As a result, many diffusion models are relegated to 2D or low-resolution volumetric applications. Incorporating text prompts, such as radiology reports, into image generation presents another challenge. Generative models utilizing text prompts as conditions demand a high-capacity denoiser to map the subtle and occasionally ambiguous pathologies mentioned in the reports to visual patterns. Enhancing the capacity of the UNet, further increases memory usage.

In 3D medical image synthesis, limitations often manifest as ``hallucinations," leading to potential biases and inaccuracies—critically concerning in medical settings. Our paper concentrates on generating Computed Tomography (CT) images of the lung. Hallucinations may manifest as missing fissure lines separating lung lobes or the creation of implausible airway and vessel structures. This problem of hallucinations is less pronounced in X-ray image generation (e.g., RoentGen~\cite{roentgen}) since the 2D X-ray projection obscures many fine details, like the lung's lobular structure, airway, and vessels. However, the issue becomes more pronounced in 3D image generation. Training on hundreds of thousands of 3D medical images isn't feasible, necessitating a strong prior to constraining the space of generative models.

To address these challenges, we propose \ourmethodN, a model tailored for high-resolution, text-guided, anatomy-aware volumetric generation. Leveraging a large dataset of approximately 9,000 3D chest CT scans paired with radiology reports, we employ a hierarchical training approach for efficient cross-modality generation. Given the input of tokenized radiology reports, our method initiates with a low-resolution synthesis ($64\times 64\times 64$), which is then fed to a super-resolution module to upscale the image to $256\times 256\times 256$. We have modified the UNet to bolster the network's capacity (i.e., the number of parameters) without significantly increasing memory requirements. \ourmethod enhances controllability by harnessing textual priors from radiology reports to guide synthesis. We further regularize the generator by creating segmentation masks of the lung's airway, vessels, and lobular structure as additional output channels alongside synthetic CT images to conserve detailed anatomical features. The resultant model can condition not only on the text but also on single or paired anatomical structures. Our algorithmic experiments and blind evaluations conducted by 10 board-certified radiologists highlight the superior generative quality and efficiency of the proposed method compared to baseline techniques. To the best of our knowledge, this is the first work to evaluate a text-to-medical image model with radiologists empirically. We also delve into the significance of components within our proposed modules.
Our code and pre-trained model is publicly available at \href{https://github.com/batmanlab/MedSyn}{https://github.com/batmanlab/MedSyn}


\section{Related Works}
\subsection{Image Synthesis Based on Text Prompt}
Text-conditional image generation enables new applications and improves the diversity of generated images compared to models that are not conditioned on text.
The most relevant studies in the domain of multi-modal generation include Imagen~\cite{IMAGEN}, Latent Diffusion (Stable Diffusion)\cite{ldm}, Video Diffusion\cite{video_diffusion}, and RoentGen~\cite{chambon2022roentgen}.
For natural images, both Imagen and Latent Diffusion pioneered the text-conditional diffusion models, producing high-fidelity 2D natural images. 
\marked{Dall·E 2~\cite{ramesh2022hierarchical} uses pre-trained CLIP model to extract features, which guides text-to-image generation}
Video Diffusion introduces a method to generate videos using score-guided sequential diffusion models, accepting text prompts as conditional inputs. 
In medical imaging domain, RoentGen enhanced the Stable Diffusion prior to achieve an exceptional generative quality for Chest X-Ray scans. 
TauPETGen~\cite{jang2023taupetgen} proposes to utilize text description as a condition for generating 2D Tau PET images.
However, these methodologies cannot be seamlessly integrated into multi-modal 3D medical volume generation because of their inherent 2D orientation or the challenges associated with efficient high-resolution 3D volume creation.
In addition, most previous methods adopt language models pre-trained on generic text, then fine-tune on it using biomedical text. In contrast, we utilize a language encoder model~\cite{boecking2022making} trained specifically with biomedical text data that learns domain-specific vocabulary and semantics.

\subsection{Generative Models for 3D Medical Imaging}
Generative models have emerged as a powerful tool in the field of medical imaging, offering a range of applications and benefits.
Previous work~\cite{kwon2019generation,xing2021cycle} leverages 3D GAN for volume generation. However, the generated images are limited to the small size of $128\times128\times128$ or below, due to insufficient memory during training. 
In order to address the challenges of producing efficient high-resolution medical volumes, HA-GAN~\cite{hagan2022} introduces an end-to-end GAN framework that operates on multiple-level feature maps. This can synthesize medical volumes of dimensions $256\times 256 \times 256$ within a single model forward pass. 
Recently, diffusion models have evolved, and both cDPM~\cite{sequential_diffusion} and SADM~\cite{sequential_diffusion2} aim to generate full 3D volumes. This sequential approach was adopted primarily because of the considerable memory demands of the 3D Attention UNet. 
In order to mitigate the discontinuity between slices due to sequential generation, MedGen3D~\cite{han2023medgen3d} introduces a diffusion refiner that generates images from three different views and averages them.
In order to address the memory constraint, Medical Diffusion~\cite{khader2022medical} proposes to use VQ-GAN~\cite{esser2021taming} to compress volume into latent space, then trains a diffusion model in the latent space.
Yet, a compromise arises between memory efficiency and speed during inference, making it challenging to scale up for high-resolution and expansive cross-modality medical volume creation.
\marked{Pinaya et al.~\cite{pinaya2022brain} have previously used LDMs to generate 3D brain MRI.}
The aforementioned limitations of previous works underscore the unique value of our proposed method for efficient high-fidelity cross-modality medical volume generation, distinguishing it from prior techniques and applications.


\section{Background}
\begin{figure*}[htp]
  \centering
    \includegraphics[width=16cm]{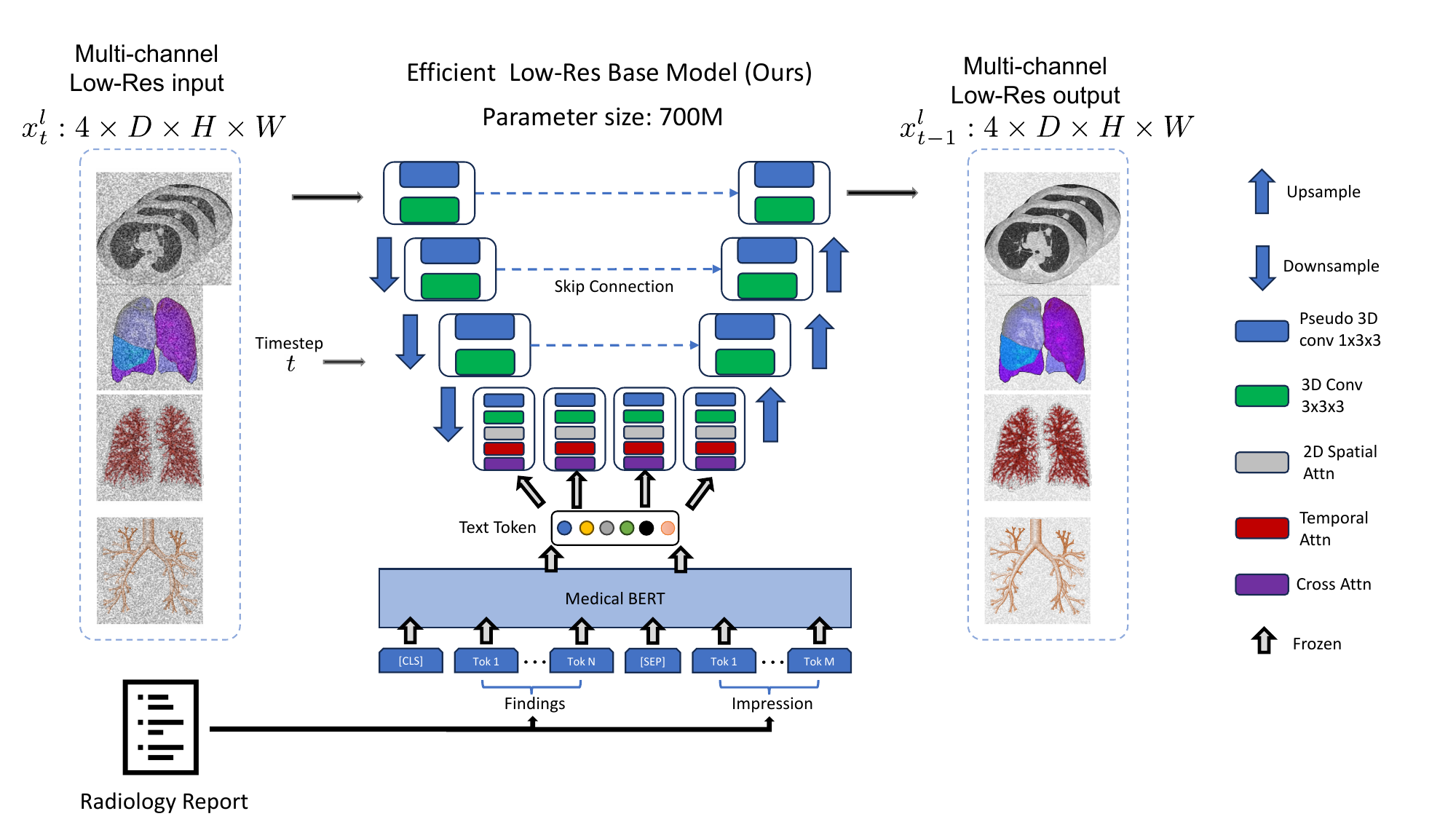}
  \caption{This figure shows our efficient low-res generative model with the clinical tokens input. In this process, we train the denoising diffusion UNet and fix the pre-trained text feature extractor of Medical BERT. To be notified, our low-res base model has a large capacity of 700 million parameters.}
  \label{fig:low_res}
\end{figure*}
Diffusion models contain two processes: a forward diffusion process and a reversion diffusion process. The forward diffusion gradually generates corrupted data from $x_0\sim q(x_0)$ via interpolating between the sampled data and the Gaussian noise as follows:
\begin{align}\label{eq:diffusion_forward}
    &q(x_{1:T}|x_0):= \prod^T_{t=1} q(x_{t}|x_{t-1}), \nonumber\\
    &q(x_t|x_{t-1}) := \mathcal{N}(x_t; \sqrt{1-\beta_t}x_{t-1}, \beta_t\textbf{I}), \nonumber
\end{align}
where $T$ denotes the maximum time steps and $\beta_t\in (0,1]$ is the variance schedule. According to Sohl-Dickstein~\textit{et al.}~\cite{sohl2015deep}, this can be reformed as a closed-form solution, from which we can sample $x_t$ at any arbitrary time step $t$ without needing to iterate through the entire Markov chain.
\begin{equation}
q(x_t|x_0)=\mathcal{N}(x_t;\sqrt{\Bar{\alpha}_t}x_0, (1-\Bar{\alpha}_t)\textbf{I}), 
\end{equation}
in which \(\Bar{\alpha}_t := \prod^t_{s=1} (1-\beta_s)\). The parameterized reversed diffusion can be formulated correspondingly:
\begin{align}
    &p_{\theta}(x_{0:T}):= p_{\theta}(x_T)\prod^T_{t=1} p_{\theta}(x_{t-1}|x_t),\\
    &p_{\theta}(x_{t-1}|x_t) := \mathcal{N}(x_{t-1}; \mu_\theta(x_t, t), \sigma_t^2\textbf{I}),
\end{align}
where we can parameterize $p_\theta(x_{t-1}|x_t)$ as Gaussian distribution when the noise addition between each adjacent step is sufficiently small, the denoised function $\mu_\theta$ produces the mean value of the adjacent predictions, while the determination of the variance $\sigma_t$ relies on $\beta_t$. The optimization objective can be written as follows:
\begin{align}\label{eq:ddpm_matching}
    \mathcal{L}=-\sum_{t>0}\mathbb{E}_{q(x_0)q(x_t|x_0)}\text{KL}(q(x_{t-1}|x_t, x_0)||p_\theta(x_{t-1}|x_t)), 
\end{align}
which indirectly maximizes the ELBO of the likelihood $p_\theta(x_0)$. When $x_0$ is given, the posterior $q(x_{t-1}|x_t,x_0)$ is Gaussian. Thus, the above objective becomes $\ell_2$ distance between the sampled $x_{t-1}$ from the posterior and the predicted denoised data\marked{, conditioning on time step $t$. The practice of predicting $x_0$ is also used in previous work~\cite{IMAGEN}.} The final minimization objective for the diffusion model \(G\) can be simplified as follows:

\begin{equation}
    \mathcal{L} = \mathbb{E}_{q(x_0)q(x_t|x_0)}\lVert G(x_t;\theta)-x_0 \rVert^2_2
\end{equation}

\section{Method}
Our radiology report conditional generated framework (\ourmethodN), as illustrated in Fig.~\ref{fig:mainSchematic}, is a hierarchical model built upon the conditional diffusion models, which take the inputs of a random Gaussian noise $\epsilon$ and the text token features$f_{text}$ to generate a $256\times 256 \times 256$ medical volumes. The proposed model has three core components: 1. A pre-trained text-encoder (Medical BERT) for extracting language features from radiology reports. 2. A text-guided low-resolution 3D diffusion model that jointly synthesizes CT volume and its anatomical structure volumes. 3. A super-resolution 3D diffusion model for scaling up the low-resolution generated volumes and complementing the missing anatomical details. In the following session, we will introduce the details of our proposed method.

\subsection{Learning Representations from Radiology Reports}
To enable the text-guided medical image generation that utilize radiology reports as the guidance, we train a text encoder (Medical BERT~\cite{boecking2022making}) that encodes natural language from radiology reports into high-dimensional vectors ($f_{text}$). The encoder schematic is shown in the lower part of Fig.~\ref{fig:low_res}. Specifically, we use the CT radiology reports in our collected data to fine-tune a BERT model~\cite{devlin2018bert} pre-trained on datasets of chest X-ray reports. We use the ``Impression", and the ``Findings" sections of CT reports as the input for the model since they provide the most characteristic descriptions related to the disease symptoms and anatomy structures. The model is then fine-tuned using two tasks: \itm{(a)} Mask language modelling; \itm{(b)} Paired section prediction. In task (a), we randomly replace $15\%$ of input tokens with $\texttt{[MASK]}$, and use the network to predict the masked token. In task (b), we randomly swap the ``Impression" section of half of the samples with another subject’s ``Impression" section. As a result, half of the samples in the dataset have unpaired ``Impression" sections and ``Findings" sections. In this way, we construct different $\texttt{[CLS]}$ embedding for paired and unpaired samples. Therefore, we train a linear classifier that uses the $\texttt{[CLS]}$ embedding to predict whether the input ``Impression" and the ``Findings" sections are paired or not. 
\subsection{Text-Conditioned Volume Generation}
This section discusses how to incorporate the text embedding from the  Medical BERT.
Direct training of high-resolution diffusion models (such as $256\times 256 \times 256$ for a given field of view) is highly memory-demanding and thus not feasible. 
We propose an efficient hierarchical model with a two-phase process. In the first phase, we generate a low-resolution volume ($64\times64\times64$) conditioned via the radiology report. In the second phase, the model outputs a high-resolution volume $256\times256\times256$ from a 3D super-solution module, which only takes low-resolution volume as input. The low-resolution image ensures the volumetric consistency of the final images. Using the radiology reports as a condition for low-res image generation is advantageous since we do not need to compromise between resolution and model capacity.
\subsubsection{Low-Resolution Volume Generation Conditioned on Reports}
To generate data conditioned via specific signals $c$, e.g., text information, we need to reformulate the unconditional DDPM objective in Equation~\ref{eq:ddpm_matching} with the conditions, 
\begin{align}\label{eq:ddpm_matching2}
    \mathcal{L}=-\sum_{t>0}\mathbb{E}_{q(x_0)q(x_t|x_0,c)}\text{KL}(q(x_{t-1}|x_t, x_0,c)||p_\theta(x_{t-1}|x_t,c)).
\end{align}
We first downsample the CT volume \(x_0\) by \(4 \times\) to produce the low-resolution one \(x_0^l\). Then, given the extracted features from the radiology report, we can finalize the training loss for the text conditional diffusion model as follows:
\begin{equation}\label{eq:low_res}
    \mathcal{L}_{low} = \mathbb{E}_{q(x_0,f_{text})q(x_t^l|x_0^l)}\lVert G(x_t, f_{text};\theta)-x_0^l \rVert^2_2,
\end{equation}
where $G$ is the denoising model with  cross-attention modules and $\theta$ is its parameters. We inject the text conditional information into the denoising network to reconstruct the original $x_0^l$ via a cross-attention mechanism. 

\subsubsection{Super Resolution Model}
Our model puts most of the learning capacity in the low-resolution module. We designed a lightweight diffusion UNet for the super-resolution module, which takes the low-resolution input and outputs the full high-resolution volume. For the super-resolution module, we design the loss to match the denoising distribution as follows:
\begin{align}
\label{eq:high_res}
    \mathcal{L}_{sup}=-\sum_{t>0}\mathbb{E}_{q(x_0,x^l_0)}&{q(x_t|x_0)}[ \nonumber\\
    &\text{KL}(q(x_{t-1}|x_t, x_0^l)||p_\theta(x_{t-1}|x_t,x_0^l))] \nonumber\\
   \mathcal{L}_{sup}^{approximate} = \mathbb{E}_{q(x_0,x^l_0)}&{q(x_t|x_0)}[ \nonumber\\
    &\lVert H(x_t, x^l_0;\phi)-x_0 \rVert^2_2],
\end{align}

where $H$ is the super-resolution denoising module with its parameters $\phi$. To be notified, we do not contain the additional text information in our super-resolution module to save the computational cost.

\subsection{Joint Generative Modeling of Volume and Anatomical Structures}
Lung CT scans exhibit more intricate details compared to other organs. These include the bronchial tree formations, the delicate vascular network, and fissure lines delineating the lung lobes. Such complexity heightens the challenge of producing high-fidelity synthetic lung CT volumes. And, those subtle anatomical structures are easily ignored during the generative modeling. Moreover, anatomical structures are usually roughly described in the clinical report, which enlarges the difficulty of CT synthesis purely based on radiology reports. Therefore, we propose to model the volume generation jointly with the anatomical structure generation, all conditioned on the radiology reports. To this end, we extract the shape information for the core anatomical structures, i.e., the lung lobes, the airway and the vessels. 
We choose the commonly used pre-trained segmentation tools to provide stable shape information for these three structures. We use lungmask~\cite{hofmanninger2020automatic} to segment lobes from CT volumes and use TotalSegmentator~\cite{wasserthal2022totalsegmentator} to segment vessels, and the NaviAirway~\cite{wang2022naviairway} is utilized for airway segmentation.
We denote the segmentation map as $l, a, v$ for the lung, airway and vessels. To enable the previous model \(G(\cdot;\theta)\) to jointly synthesize CT volume $x$ along with its structures $l, a, v$, we simply add three more channels to the input and output layers of model \(G\)\marked{, while still conditioning on time step $t$}. We then directly concatenate them in the channel dimension and construct a diffusion process on four-channel 3D volumes. To be notified, all the shape segmentations are paired with the volumes in the concatenation operation. We define the concatenation operation as $concat(\cdot)$ and the newly constructed volume as $x' = concat(x,l,a,v)$. Then, with minor modifications to Equations~\ref{eq:low_res} and~\ref{eq:high_res}, we can write down our joint generation training objectives for the low-resolution phase and high-resolution phase as follows:

\begin{equation}\label{eq:low_res_joint}
    \mathcal{L}_{low} = \mathbb{E}_{q(x'_0,text)q(x'_t|x'_0)}\lVert G(x'_t, f_{text};\theta)-x'_0 \rVert^2_2,
\end{equation}
\begin{equation}\label{eq:high_res_joint}
    \mathcal{L}_{sup} = \mathbb{E}_{q(x'_0,x'^l_0)q(x'_t|x'_0)}\lVert H(x'_t, x'^l_0;\phi)-x'_0 \rVert^2_2.
\end{equation}

\subsection{Anatomically Controllable Synthesis}
Compared with the conditional modelling of ControlNet~\cite{controlnet}, our model is significantly different, where we model the volume generation with the structure information jointly~$p(x,c)$ while ControlNet generates data conditioned on the structure~$p(x|c)$. Ours is more flexible as \ourmethod can still generate data when the predefined structures are unavailable, which is not feasible for ControlNet. Furthermore, if we marginalize one or multiple components in the joint denoising process, such as fixing the $l$ given the predefined lobe structure inputs, we can achieve exactly what ControlNet can do. Further, if we fix the $x$, we can get the structure output from our model, such as segmentation maps. We will show these advantages of our model in the experimental section. 
\subsection{Efficient 3D Attention UNet}
Although we propose an efficient two-phase text2volume generation, like other works in 3D generation~\cite{sequential_diffusion}, we still face the memory issue when generating such high-resolution volume. Sequential generation with conditional diffusion models~\cite{sequential_diffusion, sequential_diffusion2} works as one solution, but new issues will be easily introduced. Therefore, we design a new base neural architecture for much more efficient volume generation. Compared with the common 3D attention UNet~\cite{video_diffusion} for video generation, we build an encoder-decoder with pure convolutions and move all the attention mechanisms into the bottom of the UNet. In this way, we propose a more efficient base model structure that drops all the computational burden from the attention mechanism while still benefiting it from the latent space where spatial resolution is much lower. This increases the parameter size via $10\times$ more but largely increases the computational efficiency. For the super-resolution network, we almost remove all the attention modules and keep one temporal attention at the bottom of the UNet, which makes it feasible for $256\times 256\times 256$ volume inputs.

\subsection{Implementation Details}
\marked{
To pre-train our text encoder, we use 209,683 reports without paired images, and 7,728 reports with paired images as training set. We pre-train our text encoder for 5 epochs.
Our training objective for the diffusion models is a simple reconstruction loss: the $\ell_2$ pixel-level reconstruction between the ground truth and the prediction for different denoising time steps without any other terms, such as perceptual loss~\cite{lin2023diffusion}.
}
Following the common choice of training diffusion model, we use a continuous cosine time scheduler~\cite{ddpm}. 
\marked{For the training of both the low-resolution diffusion model and the super-resolution model, the number of time steps is set as 1000. During inference, we use 50 DDIM steps for the low-resolution diffusion model, and 20 DDIM steps for the super-resolution model.}
For the optimizer, we apply AdamW~\cite{adamw} with the learning rate of $1\times 10^{-4}$ and $\beta=\{0.9, 0.999\}$ with clipping the gradient norm bigger than 1. We apply the mixed-precision for optimizing models to make the training more efficient. The gradient accumulation is applied during training to scale up the training batch size, as diffusion models are sensitive to the small training batch size~\cite{ghalebikesabi2023differentially}. Ultimately, we can train our 700M parameters low-resolution base model with a batch size of 64 and the super-resolution model with a batch size 32 on four NVIDIA A6000 GPUs. 
Both models have converged after undergoing 40k iterations of training.

\section{Experiments}

\begin{figure*}[htp]
  \centering
    \includegraphics[width=\textwidth]{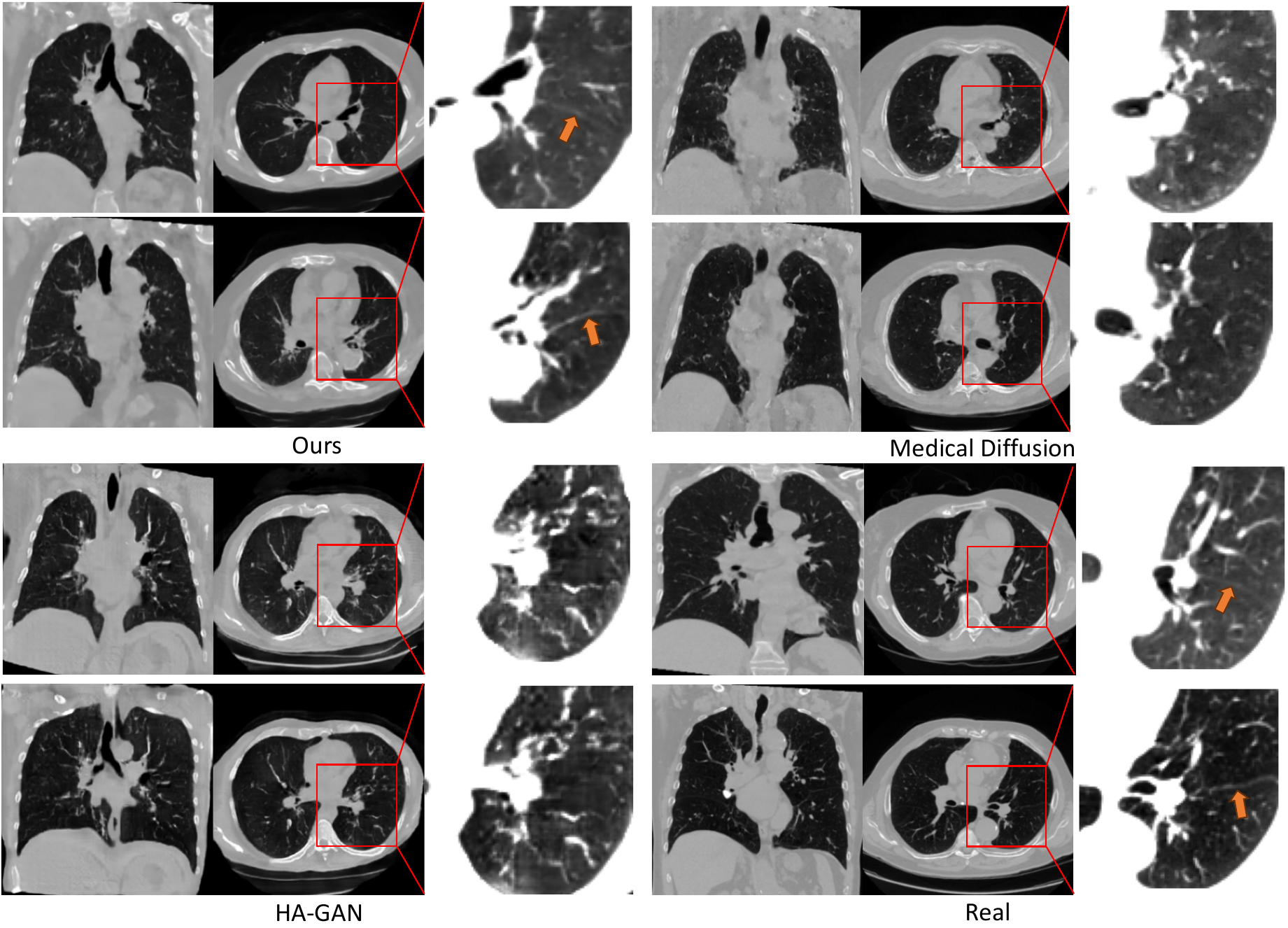}
  \caption{Randomly generated images (from HA-GAN and Medical Diffusion) and the real images. The first two columns show axial and coronal slices, which use the HU range of [-1024, 600]. The last column shows the zoom-in region and uses HU range of [-1024, -250] to highlight the lung details. Our method is the only one that can preserve delicate anatomical details, including fissures, as indicated by the arrows.}
  \label{fig:result_synthesis}
\end{figure*}
\begin{figure*}[htp]
  \centering
    \includegraphics[width=18cm]{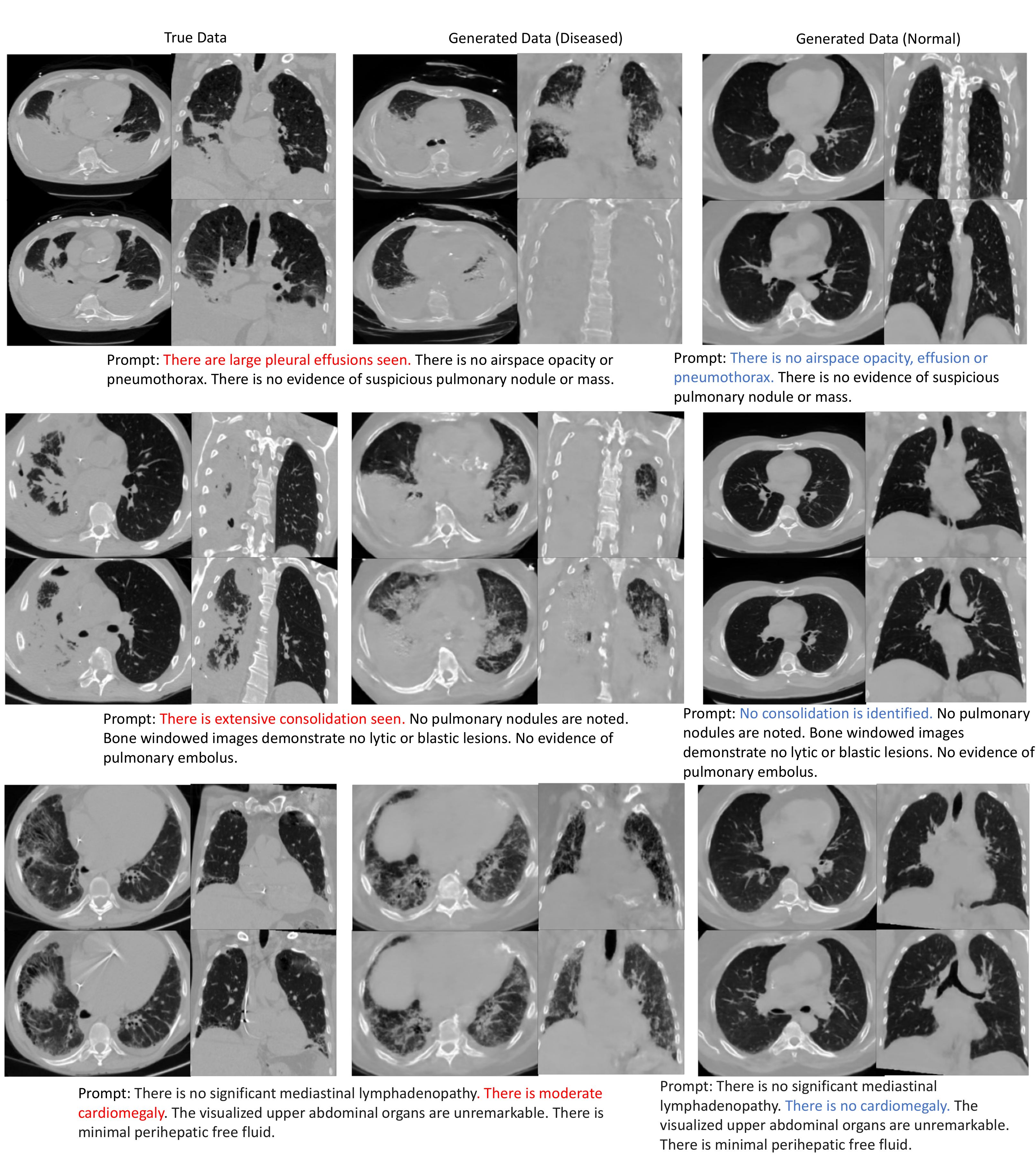}
  \caption{Images conditionally generated with disease-related prompts. We show the real images in the first two columns.
  Then we extract disease-related mentions from their associated reports as conditions to generate images, which are shown in the third and fourth columns. 
  We also show the synthesized samples by conditioning on prompts reversed of the disease in the last two columns. Four slices are shown for each subject. \marked{The generated images are conditioned on text only.}}
  \label{fig:prompt_comparison}
\end{figure*}
This section presents a comprehensive evaluation of the proposed generative model, \ourmethodN. We will first describe the dataset used in this experiment. Then, the \ourmethod is compared with the state-of-the-art GANs and diffusion model, including WGAN~\cite{gulrajani2017improved}, $\alpha$-GAN~\cite{kwon2019generation}, HA-GAN~\cite{hagan2022} and Medical Diffusion~\cite{medicaldiffusion}. Extensive comparisons and analysis are finally given to evaluate the effectiveness of our method, qualitatively and quantitatively.
\marked{For training of all baseline methods, we use the author's implementation. We made minimal modifications to the code to adapt to our dataset.
All models are trained from scratch, for fair comparison with our method.}

\subsection{Dataset}
We conduct experiments on a large-scale 3D dataset, which contains
3D thorax computerized tomography (CT) images and associated radiology reports from 8,752 subjects.
The dataset also contains 209,683 reports without corresponding images.
The images and reports are collected by the University of Pittsburgh Medical Center and have been de-identified.
\marked{
We randomly split our dataset of 8,752 subjects into a training set consisting of 7,728 subjects (88\%), and a validation set of 1,024 subjects (12\%).
}
The images have been pre-aligned using affine registration and re-sampled to $1 mm^3$.
We resize the images to $256\times256\times256$. We use the nearest-neighbor downsampling to reduce the scans \(4\times\) to train the low-resolution base model. The Hounsfield units of the CT images have been calibrated, and air density correction has been applied. The Hounsfield Units (HU) are mapped to the intensity window of $[-1024,600]$ and normalized to $[-1,1]$.

\subsection{Evaluation for Synthesis Quality}

\begin{table}[htp]
 \centering
 \caption{ Quantitative comparison with different methods. Our method outperforms baseline methods in terms of distance metrics, and preserves airway better}
 \label{tbl:FID}
  \begin{tabular}{lccc}
   \toprule
    Method &  FID$\downarrow$&  MMD$\downarrow$ & Airway ($\times 10^4 mm^3$)$\uparrow$\\
   \midrule
   WGAN & $0.070$ & $0.094$ & $1.07_{\pm0.64}$\\
   $\alpha$-GAN& $0.028$& $0.057$& $1.14_{\pm0.68}$ \\
   HA-GAN & $0.023$ & $0.054$ &$2.04_{\pm0.73}$\\
   Medical Diffusion & $0.013$ & $0.022$ & $1.77_{\pm0.93}$\\
    \midrule
 \textbf{Ours}  & $\bm{0.009}$ & $\bm{0.019}$ & \bm{$3.34_{\pm1.19}$}  \\
  \textbf{Ours} w/o shape & -& -& $1.99_{\pm1.05}$ \\
   \midrule
   Real & -&- & $4.58_{\pm1.45}$ \\
  \bottomrule
  
  \end{tabular}
  \vspace{-3mm}
 \end{table}

\subsubsection{Quantitative Evaluation}
If the synthetic images are realistic, then their distribution should be indistinguishable from that of the real images.
Therefore, we can quantitatively evaluate the quality of the synthetic images by measuring the distance to the real data, using Fréchet Inception Distance (FID)~\cite{heusel2017gans} and Maximum Mean Discrepancy (MMD)~\cite{gretton2012kernel}. The lower the FID/MMD value is, the more similar the synthetic images are to the real ones. We use a sample size of 1,024 for computing the FID and MMD scores.
For our method, we use randomly selected reports as the condition.
To compute FID and MMD scores for 3D CT scans, like \cite{hagan2022}, we leverage a pre-trained 3D ResNet model on medical images~\cite{chen2019med3d} for feature extraction. 
As shown in Table~\ref{tbl:FID}, our method achieves lowest FID and MMD than the baselines, which implies that our diffusion model generates more realistic images.

\noindent \textbf{Quantitative Evaluation on Anatomical Details:}
While metrics like FID and MMD are widely used in literature and empirically work well for natural images, they highlight the semantic-wise similarly (distance) but may ignore subtle but important anatomical details in medical images, as implied by the small (FID/MMD) gap between different methods. Their real distances, as later shows in Fig.~\ref{fig:result_synthesis}, could be much bigger when taking account into the anatomical details we are focused on. Therefore, we evaluate how well the generated images can preserve the anatomical details. Specifically, we use Total Segmentor~\cite{doi:10.1148/ryai.230024} to segment vessels and airways from generated images and real images, and measure the volume. The results are shown in Table.~\ref{tbl:FID}.
We also perform statistical tests (one-tailed two-sample \emph{t}-test) on the evaluation results. At the significance level of $p<0.05$, the results are significant for all three conditions, which further identify the effectiveness of our model on prompting generation.

\subsubsection{Qualitative Evaluation}
To qualitatively analyze the results, we show cases of synthetic images from current state-of-the-art GAN~\cite{hagan2022} and diffusion model~\cite{khader2022medical}. As in Fig.~\ref{fig:result_synthesis}, although synthetic images from different methods are all closed to the real ones in overall appearance, only our \ourmethod consistently produces anatomically plausible CT scans upon closer inspection, showcasing its superiority.

\subsection{Evaluation for Conditional Generation}
\begin{table}[htp]
 \centering
 \caption{ Evaluation for conditional generation of pleural effusion. \marked{We measure the segmented volume of pleural effusion from generated images conditioned on different prompts}}
 \label{tbl:cond_effusion}
  \begin{tabular}{lc}
   \toprule
    Prompt type &  Pleural effusion volume (L) \\
   \midrule
   No effusion & $0.00_{\pm.00}$ \\
   Large effusion & $1.73_{\pm.22}$  \\
  \bottomrule
  \end{tabular}
 \end{table}
 \begin{table}[htp]
 \parbox{.45\linewidth}{
 \centering
 \caption{ Evaluation for conditional generation of bullous emphysema. The results show that the bullae mentioning can increase the emphysema volume in generated volumes }
 \label{tbl:cond_bullae}
  \begin{tabular}{lc}
   \toprule
    Prompt type &  \%LAA-950 \\
   \midrule
   No bullae & $0.019_{\pm.018}$  \\
   With bullae & $1.4_{\pm3.5}$ \\
  \bottomrule
  \end{tabular}}
\parbox{.45\linewidth}{
  \centering
 \caption{ Evaluation for conditional generation of cardiomegaly. The results show that the cardiomegaly mentioning can increase the heart size in generated volumes }
 \label{tbl:cond_cardiomegaly}
  \begin{tabular}{lc}
   \toprule
    Prompt type &  CTR \\
   \midrule
   No cardiomegaly & $0.48_{\pm.06}$  \\
   With cardiomegaly & $0.75_{\pm.24}$ \\
  \bottomrule
  \end{tabular}}
 \end{table}
\begin{figure}[htp]
  \centering
    \includegraphics[width=9cm]{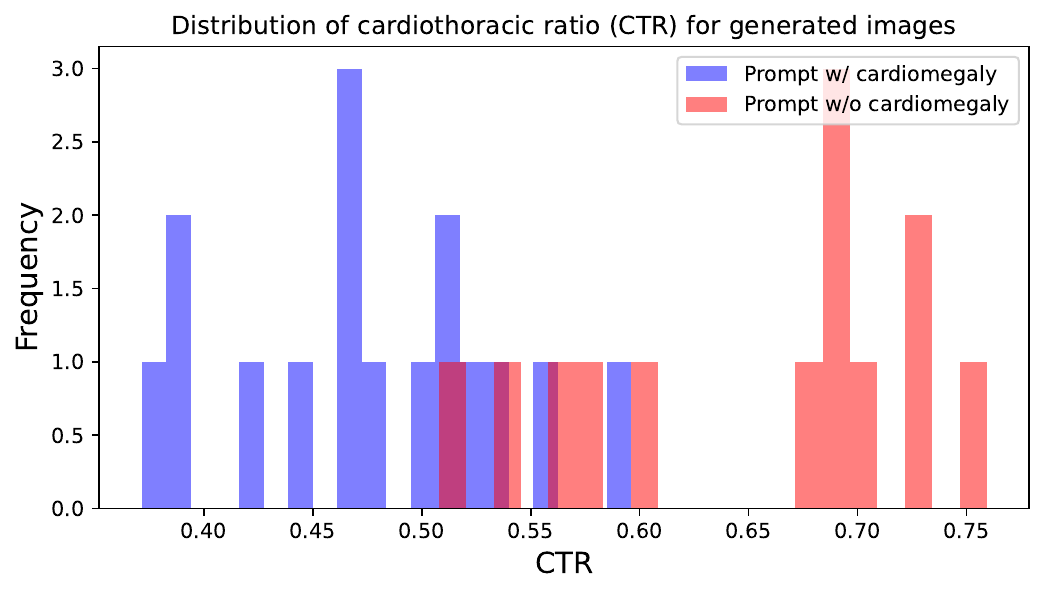}
  \caption{Distribution of cardiothoracic ratio for images generated conditioning on different prompt types. The results show that when feeding prompt with cardiomegaly mentioning, the generated images will have higher CTR.}
  \label{fig:histogram_CTR}
\end{figure}
\begin{figure*}[htp]
  \centering
    \includegraphics[width=17cm]{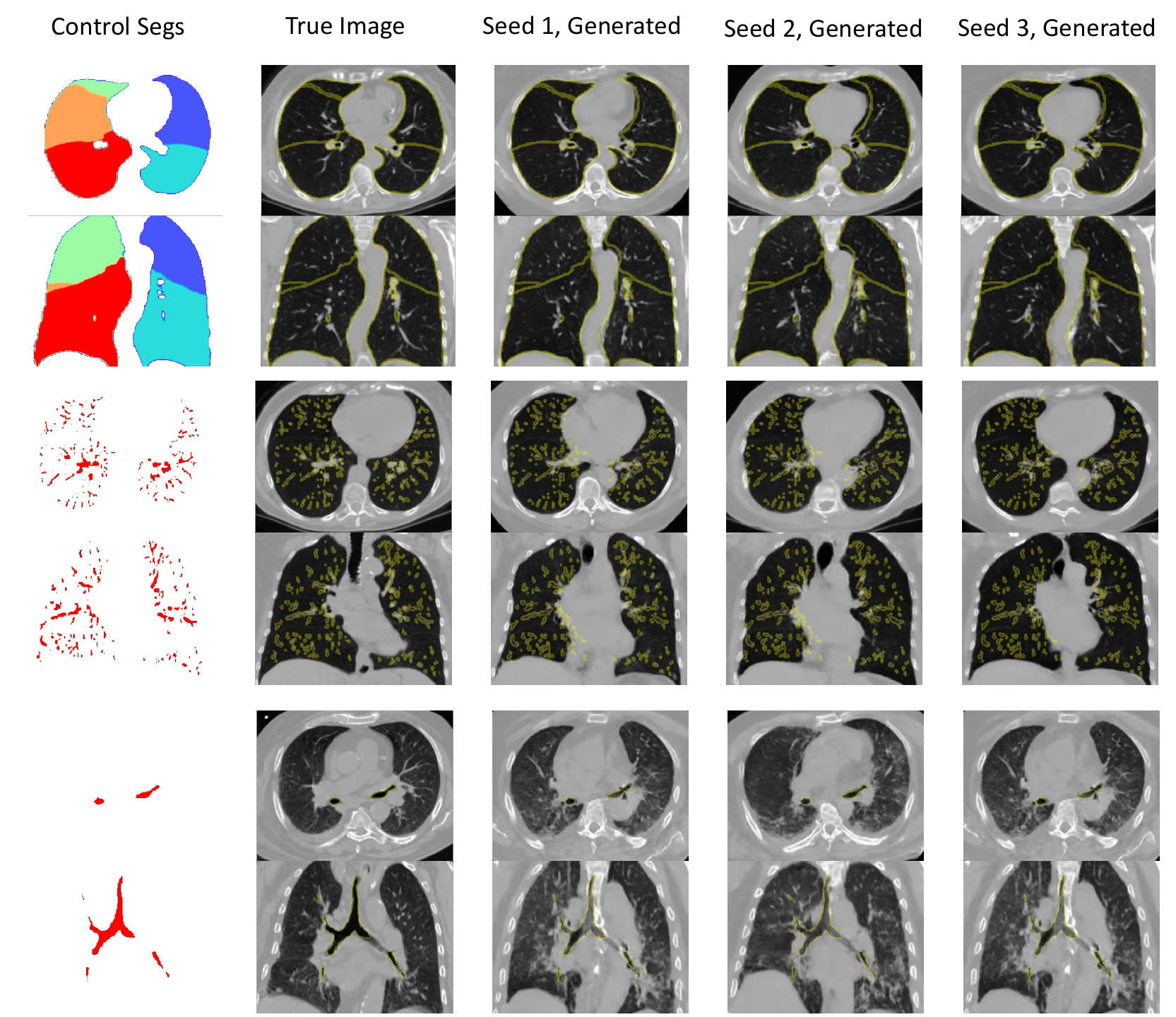}
  \caption{Controlled volume synthesis via the anatomical priors. The first column shows the anatomical mask used as the condition. The second column shows the corresponding real images. The remaining columns show samples of conditionally generated images. The results show that the generated images can preserve the conditioning anatomical structures.}
  \label{fig:marginalization}
\end{figure*}
In this section, we perform experiments to study the relevance of generated images in response to specific input prompts, specifically on pleural effusion, bullous emphysema, and cardiomegaly.
First, we build prompt pairs by selecting prompts from real reports that contain descriptions about certain pathology (e.g. \emph{There are large pleural effusions seen}), and reversing its description (e.g. \emph{There is no pleural effusion}) to build its prompt pair.  The prompts used here can be found in Fig.~\ref{fig:prompt_comparison}. Then, we use our model to generate images conditioning on the original prompts and the modified prompts, respectively. 
Conditioned on each prompt, we generate 32 CT volumes and perform quantitative analysis to measure the alignment between the synthetic images and the abnormality condition specified in the prompt. For pleural effusion, we use Total Segmentor~\cite{doi:10.1148/ryai.230024} to segment the effusion from generated images and measure the volume. For bullous emphysema, we measure the \%LAA-950 (percentage of low attenuation areas less than a threshold of -950 Hounsfield units) for generated images. For cardiomegaly, we use Total Segmentor~\cite{doi:10.1148/ryai.230024} to segment the heart and lung region from CT volume, and then we compute the cardiothoracic ratio (CTR) by measurement of the maximal cardiac width divided by the maximal thoracic width at the same axial scan level. The evaluation results for pleural effusion, bullous emphysema and cardiomegaly are shown in Table.~\ref{tbl:cond_effusion}, ~\ref{tbl:cond_bullae} and ~\ref{tbl:cond_cardiomegaly} respectively. For pleural effusion, we found that when conditioning on the prompt with ``large effusion,'' the generated images show a greater volume of pleural effusion compared to images synthesized with a prompt containing ``no effusion.'' For bullous emphysema, we found that the generated images conditioning on prompt containing ``no bullae" have higher \%LAA-950 values, which suggests more severe emphysema. For cardiomegaly, we found that when conditioning on the prompt with ``There is cardiomegaly,'' the generated images have higher CTR, which suggests a greater degree of cardiomegaly. We also provide the distribution of CTR in Fig.~\ref{fig:histogram_CTR}.

For the qualitative examples generated from our model, we chose those three distinct prompts paired with negative prompts to show the prompting effect on synthetic images. In Fig.~\ref{fig:prompt_comparison}, we show the volumes from the real and synthetic data with the text description and the negative prompting synthetic data. Our model shows the ability to generate unseen data and control the generative process through prompting.

\subsection{Controllable Synthesis via Conditioning on Anatomical Structures}
In this section, we explore the application of conditional generation. In this study, we aim to generate data when the anatomical structures are available, such as we can simulate the structures of lobes, airways or vessels. Conditioned on those priors, we are able to provide volumetric CT scans through our model. In Fig~\ref{fig:marginalization}, we fix the lobes, vessels or the airway in the input channel, respectively, which are segmented from the real data. Then, we generate the lungs with those anatomical structures, which shows great consistency with those anatomical priors and variance for different seeds.

To be notified, if we marginalize the input channel of the volumes, we can get the segmentation from the rest of the channels, which means our model can perform the segmentation task as well. However, the generative quality is mostly dominated by the low-res module, thus, the segmentation cannot be compared with the state of the art. To show the potential of future multimodal segmentation via prompt, we compare the segmentation quality of the model with prompt or without the prompt (by setting zero values for the conditional text embedding).
In our experiments, we use the embedding extracted from the associated radiology report as the condition. The evaluation results are shown in Table.~\ref{tbl:cond_gen_mask}. \marked{We use Dice score as the evaluation metric for segmentation. We note that the baseline methods can't use segmentation as conditioning, so we don't include the results here.} We find that our method achieves decent segmentation performance. With additional information from reports, our method performs even better. In this sense, we show that our generative model has the potential for prompt-guided volumetric segmentation.
\begin{table}[H]
 \centering
 \caption{ Evaluation for segmentation. With additional
information from reports, the segmentation is further improved}
 \label{tbl:cond_gen_mask}
  \begin{tabular}{lcc}
   \toprule
    \marked{Dice score}$\uparrow$ &  Airway & Lobe \\
   \midrule
   No text prompt & $0.70_{\pm.14}$  & $0.69_{\pm.12}$\\
   With text prompt & $0.75_{\pm.10}$ & $0.77_{\pm.12}$\\
  \bottomrule
  \end{tabular}
 \end{table}

\subsection{Data Augmentation for Supervised Learning}
\marked{In this experiment, we used the synthesized samples from MedSyn to augment the training dataset for a supervised classification task.
Previous work~\cite{frid2018gan} has shown that synthetic samples improve the diversity of the training dataset, resulting in a better discriminative performance of the classifier.
Motivated by their results, we conduct new experiments of using our MedSyn for data augmentation.
First, we train a classifier for predicting lung opacity and pleural effusion from CT scans in RADChest dataset~\cite{draelos2021machine}. Second, we use the SARLE labeler~\cite{draelos2021machine} to parse the reports in the validation set of our UPMC dataset to derive the labels for lung opacity and pleural effusion. Next, we randomly sampled 100 reports parsed for positive and negative for the two diseases mentioned above. Then, we use the 200 reports as the condition, and feed them into our MedSyn model to generate additional samples. Finally, we use the synthesized samples as extra samples to train the classification models. We perform 5-fold cross-validation. The results are shown in Table.~\ref{tbl:data_aug}. We found that when using augmented samples from our MedSyn, the performance of the classifier can be improved.}

\begin{table}[htp]
\caption{\marked{Evaluation for data augmentation. Baseline model is trained only with real RADChest data. We also augment the training set with 200 MedSyn-generated samples, and report the accuracy and F1 score.}}
\centering
\begin{adjustbox}{max width=0.5\textwidth}
\centering
\label{tbl:data_aug}
 {\color{mymarked}\begin{tabular}{lcccc}
   \toprule
  Method &\multicolumn{2}{c}{Pleural effusion}&\multicolumn{2}{c}{Lung opacity}\\
  \toprule
Metric  &Accuracy\%$\uparrow$&F1$\uparrow$&Accuracy\%$\uparrow$&F1$\uparrow$\\
  \midrule
Baseline &
$90.7_{\pm3.2}$&$0.79_{\pm.05}$&
$61.0_{\pm2.2}$&$0.72_{\pm.03}$\\
Augmented w/ MedSyn &\bm{$94.0_{\pm.2}$}&\bm{$0.84_{\pm.01}$}&
\bm{$62.0_{\pm1.5}$}&\bm{$0.75_{\pm.00}$}\\
 \bottomrule
 \end{tabular}}
\vspace{-3mm}
\end{adjustbox}
\end{table}

\subsection{Evaluation by Radiologists}
\label{sec:user_study}
\marked{To complement the evaluations of our method, we designed a blind evaluation survey that elicits board-certified radiologists' opinions on the anatomical feasibility of structures generated by our approach in comparison with those generated by existing methods, including Medical Diffusion~\cite{medicaldiffusion} and HA-GAN~\cite{hagan2022}, and real CT scans. The survey also exposes how accurately radiologists can recognize pathologies in CT scans generated by pathology prompts using our method. To achieve a blind evaluation of the generated CT scans, we intentionally did not mention to the radiologists that some of the CT scans they were about to interpret were generated by AI to avoid potentially biased results due to potentially negative perceptions towards AI.}

\marked{Overall, our findings from the survey with 10 radiologists with varying years of experience (4 -- 23 years) reveal that radiologists can correctly recognize the pathologies defined by the prompt in the CT scans generated by our method with high accuracy. Additionally, our findings reveal that our method generates CT scans with fissure lines and lobe structures that are significantly more anatomically feasible than in CT scans generated by the Medical Diffusion~\cite{medicaldiffusion} and HA-GAN~\cite{hagan2022} methods. Our findings also reveal that our methods generate CT scans with airway structures that are significantly indistinguishable from real CT scans and more anatomically feasible than those in CT scans generated by the Medical Diffusion method. We expand upon the experiment design that led to these findings and additional statistical analyses below.}

\subsubsection{Experiment Design}

\marked{We designed an online survey to elicit radiologists' opinions on how accurately our method represents different diseases and how our method compares to generated CT images from related works and real CT images. To explore this, our survey consisted of pathology recognition and anatomical feasibility evaluation tasks. We designed an interface for both tasks to simulate how radiologists view CT scans by showing them the axial, coronal, and sagittal views. Our interface, built on a medical research image viewer called Papaya~\footnote{\href{https://github.com/rii-mango/Papaya}{https://github.com/rii-mango/Papaya}}, is embedded into a Qualtrics survey. We chose to use Papaya due to the limitations of Qualtrics, and professional viewers such as ITKSnap or 3DSlicer cannot be embedded into an anonymized survey platform like Qualtrics. Hence, we have limitations for the contrast window that radiologists use when viewing vessels. Therefore, the evaluation of the vessels has limitations in terms of visualization. Our interface did allow radiologists to adjust the contrast of the images and swap the main image with one of the other two views.}

\marked{All radiologists were provided an instruction video to show them how to interact with the interface before they started the survey. During the instruction video, all radiologists were told that they would review CT scans ``that may belong to different patients and were acquired based on different image acquisition devices and image reconstruction methods''.  Additionally, one board-certified radiologist and one of the authors were available while the participant took the survey to help address any confusion or technical issues.}

\marked{We recruited 10 board-certified radiologists with varying years of experience to participate in our study. All radiologists were recruited through our professional network. }

\marked{\paragraph{Pathology Recognition Task}
The pathology recognition task in the survey asks radiologists to identify the most prominent finding in the CT scan from a selection of options: consolidation, pleural effusion, cardiomegaly, no abnormalities, and other abnormalities not listed. They could leave a note in an optional text field if they feel multiple prominent or other findings are present. Radiologists are shown six CT scans in total: five generated CT scans by our method and one real CT scan. We generated five CT scans, including one for cardiomegaly, one for consolidation, two for pleural effusion, and one for no abnormalities. The real CT scan shown is randomly selected to present one of those conditions. The six CT scans are shown to the radiologist in a random order.}

\marked{\paragraph{Anatomical Feasibility Task}
The anatomical feasibility task specifically asks radiologists to rank the four CT scans based on how well they preserve the given anatomical structure, considering which most looks like it is of a real image where $1$ is best preserved and $4$ is least preserved. This task evaluates the realism of three categories of anatomical structures: the lobe structure along with the fissure lines, the vessel structures, and the airway structures. For each category, we show the radiologists four different sets of four CT scans to rank. The four sets and the four CT scans for a given set are randomized in the order they appear. Lastly, the anatomical structure to be identified is shown in a randomized order.}

\subsubsection{Analysis}

\marked{
For the pathology recognition task, we calculate the total number of radiologists that correctly identify the pathology used in the prompt to generate that CT scan. We cross-check the radiologists' open-ended responses for correctness with a radiologist within our professional network.}

\marked{For the anatomical feasibility task, we collect four data points from each participant for the rank of each method for the three categories (lobe structure along with lung fissures, vessel structure, and airway structure). With this ranked data, we calculate the frequency of participants that ranked each method for each rank and calculate the mean rank. We use a non-parametric Friedman test to determine if the mean rank for each method is significantly different from that of the other. We conduct a pairwise analysis to determine which mean ranks are significantly different from each other and use Bonferonni Correction to adjust the p-values due to multiple comparisons. We set our significance threshold as $p<= 0.05$. We calculate Kendall's Coefficient of Concordance to determine how consistent the rankings are across the radiologists.}

\subsubsection{Findings}

\marked{The radiologists who completed the survey included four senior-level radiologists with more than 15 years of experience in radiology and six junior-level radiologists with four to eight years of experience in radiology. One of the radiologists is currently in their residency, and another is a faculty member. The survey took the radiologists $37.47$ minutes on average to complete ($std=12.06$ minutes). Below are additional statistics on each task.}

\begin{figure}[!h]
  \centering
    \includegraphics[trim={1cm 1cm 1cm 1cm}, width=8cm]{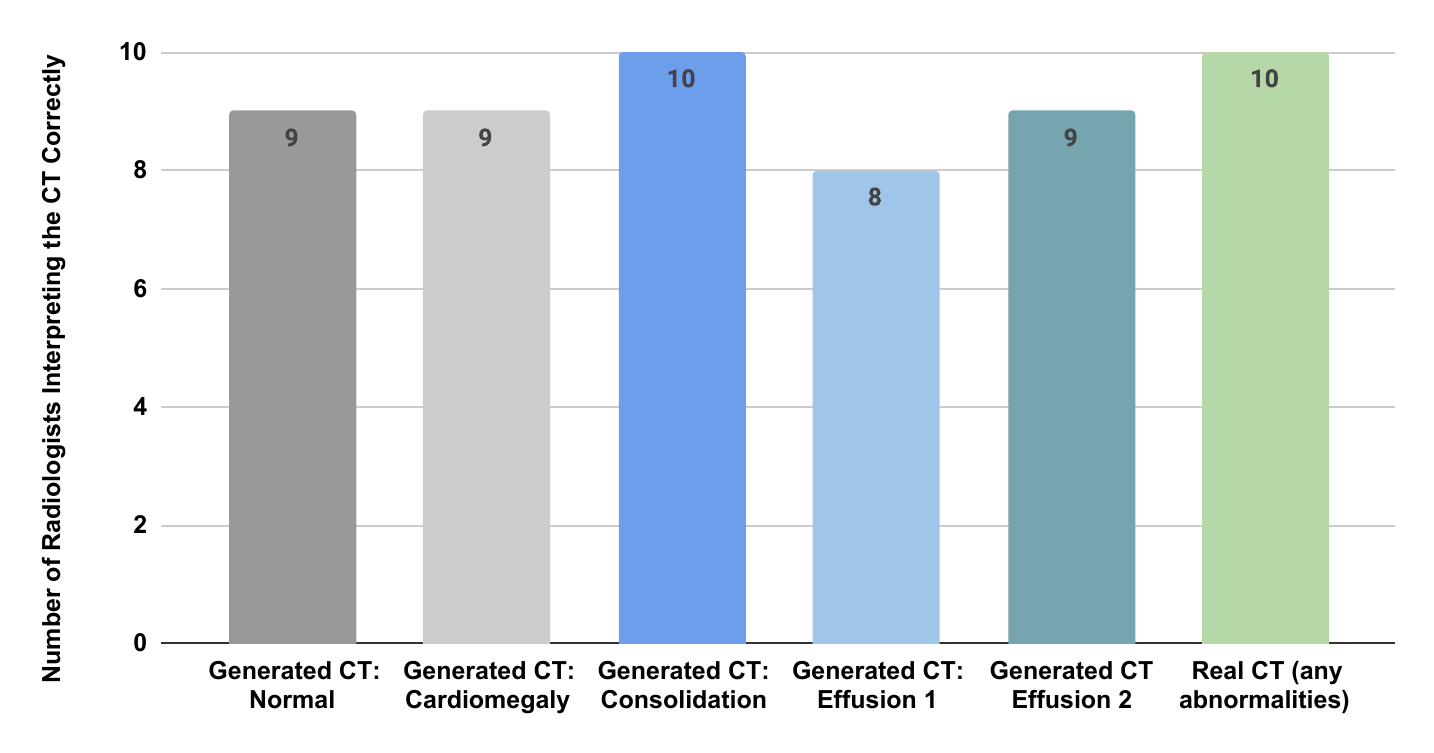}
  \caption{\marked{The number of radiologists that correctly interpreted the CT scans generated by our method for different conditions and real CT scans for different conditions.}}
  \label{fig:diagnostic-task}
\end{figure}

\marked{\paragraph{Pathology Recognition Task} Figure \ref{fig:diagnostic-task} shows that every radiologist's interpretation was consistent with the pathology of the real CT scan and the prompted pathology in the generated CT scan representing consolidation. Nine of ten radiologists correctly interpreted the pathology for the CT scans generated with a pathology prompt for cardiomegaly, pericardial effusion, and normal conditions. Eight out of 10 radiologists correctly recognized the other pericardial effusion case. One radiologist interpreted this CT scan to present consolidation, while another mentioned another abnormality that was not listed.}

\begin{figure}[!h]
  \centering
    \includegraphics[trim={3cm 3cm 15cm 0cm}, width=8cm]{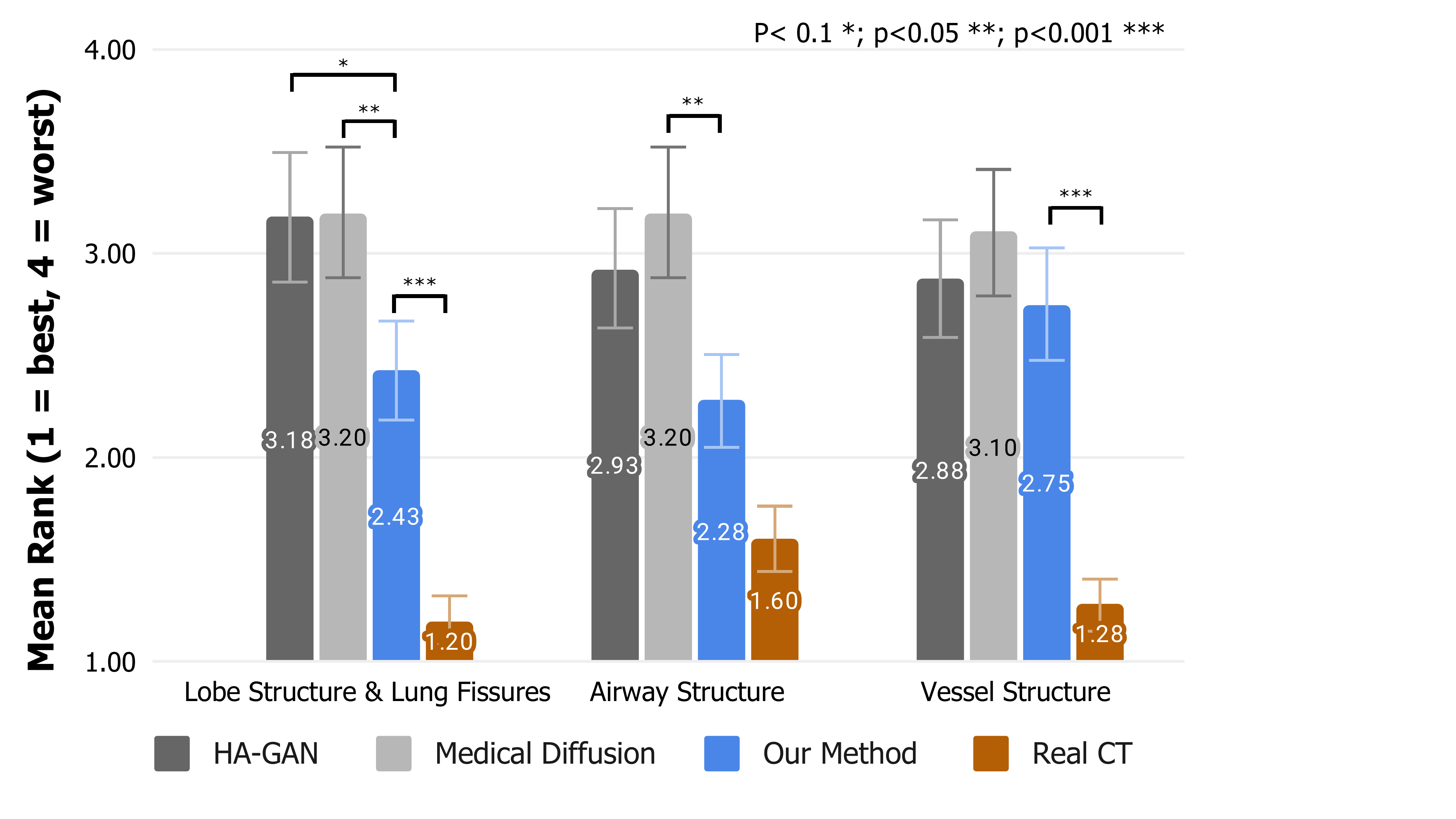}
  \caption{\marked{The mean ranks for each method and each structure. We show the p-values for our method compared to the other methods and the real CT images. P-values for comparisons that do not include our method are not included. }}
  \label{fig:means}
\end{figure}

\marked{\paragraph{Anatomical Feasibility Task}
We found that the CT scans generated by our method present lobe structures and lung fissures that are significantly more anatomically feasible than those generated by the Medical Diffusion $(p < .05)$ and weakly significantly better than CT scans generated by the HA-GAN $(p=.056)$. As seen in Figure \ref{fig:means}, the real CT scans are significantly better than all three of these methods, as expected. Research images achieve higher quality than clinical images because of limited data and limited exposure patients can have with the imaging machines. The W from Kendall's Coefficient Test is $0.528$, indicating moderate agreement among the rankings provided by the radiologists for this structure $(p < .001)$, validating the effect of these findings.}

\marked{For the airway structures, CT scans generated by our method produce airway structures that are significantly more anatomically feasible than those generated by the Medical Diffusion method $(p<.05)$. Interestingly, our method is able to generate the airway structures in CT scans to be significantly indistinguishable from those in real CT scans $(p=0.1)$. The radiologists, on average, ranked the anatomical feasibility of the airway structures in the real CT images as $1.6$ and $2.28$ for our method (see Figure \ref{fig:means}).}

\marked{For the vessel structure, the real CT scans are significantly better than all three methods $(p=0.0)$, and our method is indistinguishable from the Medical Diffusion and HA-GAN $(p=1.0);$ (Kendall's Coefficient $W=0.413, p=<.001$). We hypothesize that we do not see any significance for the vessel structure because we train our MedSyn with masks predicted by pre-trained segmentation models, and those segmentation methods are more reliable for the airway (NaviAirway~\cite{wang2022naviairway}) than the vessel structures (TotalSegmentator~\cite{doi:10.1148/ryai.230024}). 
Additionally, the contrast of the CT scan was not pre-determined for the radiologists for this structure, and some chose not to adjust the contrast.}

\subsection{Ablation study on the language model}
\marked{In our MedSyn model, we leverage a language model pre-trained on biomedical text to extract conditioning features for the diffusion model.
In this section, we conducted new ablation experiments to demonstrate the effectiveness of our biomedical language model.
Specifically, we adopt BERT-Base~\cite{devlin2018bert} as pre-trained language model, which is pre-trained on general text corpora, including English Wikipedia and BookCorpus dataset~\cite{Zhu_2015_ICCV}. Next, we fine-tuned the model with our radiology report dataset. Finally, we re-train our diffusion model with the standard language model, and perform the evaluation.
 The results are shown in Table.~\ref{tbl:cond_effusion}. It's evident that our choice of using a language model specifically pre-trained on biomedical text improves the sensitivity of the diffusion model to text prompts.}
 
 \begin{table}[htp]
 \centering
 \caption{\marked{Evaluation for conditional generation of pleural effusion. Our biomedical language model is more sensitive to the effusion mentioning in the prompt}}
 \begin{adjustbox}{max width=.5\textwidth}
 \label{tbl:cond_effusion}
  {\color{mymarked}\begin{tabular}{lcc}
   \toprule
    Measured effusion volume (mL) & Generic pre-trained LM & Biomedical pre-trained LM\\
   \midrule
   Prompting w/o effusion& $0.6_{\pm1.4}$ & $0.0_{\pm0.0}$ \\
   Prompting w/ effusion& $2.4_{\pm5.9}$ & $1725.7_{\pm219.5}$  \\
  \bottomrule
  \end{tabular}}
  \end{adjustbox}
 \end{table}

\section{Discussion}
In this study, we achieve high-fidelity, anatomy-aware synthesis of volumetric lung CT scans using guidance from radiology reports. Nonetheless, our model has certain limitations. First, the anatomical structures of the lobe, airway, and vessel are derived from pre-trained segmentation networks. This method may not always align with the ground truth, particularly concerning detailed airway and vessel structures. 
Second, the radiology reports only provide a condition on the low-resolution images. Therefore, if the text condition mentions subtle changes that require high-resolution, the model will likely be unable to generate that. In other words, our current model is good at generating global and large-scale changes. 
While our approach demonstrates remarkable adaptability in generating volumetric lung CT scans from radiology reports, evaluating more intricate lung diseases remains challenging due to the complexities presented in the reports. Such evaluations might best be deferred to radiology experts. \marked{Therefore, we conduct a blind user study by radiologists in Section.~\ref{sec:user_study}, which verified the quality and fidelity of the generated images.} 
\marked{During the inference time, our models support both generation with only one conditioning type, and generation with both text and shape conditioning. 
 If the user thinks there might be a conflict between the text prompt and anatomical shape conditioning, he/she can choose to use one conditioning.}

The existing conditional generative models utilized in medical imaging have limited capabilities. They can either accommodate discrete conditions (such as the presence and absence of a disease) or are limited to only 2D images (i.e., X-ray images) when conditioned on text. Utilizing free-style text as a condition can yield substantial enhancements in the diversity of the generated samples. One can control the pathology and the anatomical location, severity, size, and many other aspects of the pathology. This technique could mitigate the longstanding issue of releasing extensive medical imaging datasets. For example, collecting datasets for rare pathology is challenging. Synthetic samples from a well-trained generative model with a radiology report can be viewed as the second-best replacement in such a scenario.  
While our approach cannot entirely substitute the release of the real data, collaborative efforts within the medical imaging community to refine this model on diverse datasets and share it can significantly mitigate this issue in the foreseeable future.

\marked{Our goal is to enhance the memory efficiency of the diffusion model without compromising its fidelity. Memory demand poses a significant bottleneck for generating high-resolution 3D medical images. To address this challenge, we introduce a two-stage stacked diffusion model. Unlike latent diffusion models (LDM), our hierarchical approach conducts denoising operations directly on image pixels, offering finer control over image generation and preserving higher fidelity details.
Our hierarchical diffusion model incurs higher computational costs when compared to LDMs. There exists a natural trade-off between fidelity and computational efficiency. However, in our scenario, the advantages of heightened fidelity surpass the disadvantages of slower generation times. Recently, there have been advancements in methods~\cite{liu2023instaflow, xu2023semi} that markedly decrease the necessary denoising steps, sometimes even to just one step. We anticipate that integrating these methods could narrow the disparity between our approach and LDMs, and we leave this as a prospect for our future work.}

The proposed model has several fascinating applications that could be pursued in the future. Data augmentation is the most obvious application of the conditional generative models~\cite{chen2021generative,shin2018medical}. One can use the generative model as a building block for model explanation, as suggested in~\cite{singla2023explaining,montenegro2021privacy,mauri2022accurate}. Generated samples can be used to audit the uncertainty of pre-trained Deep Learning models by conditioning on the pathology, changing various aspects of the anatomy, and assessing the DL model's output distribution. One can deploy such an approach to evaluate and improve the out-of-sample distribution of the DL model for various tasks such as classification and segmentation~\cite{mahapatra2021interpretability,li2021semantic,yan2019domain,xie2020mi}. Since our model can condition anatomical segmentation and generate a consistent volumetric image, one can use synthetic data to train a data-free and robust segmentation method similar to~\cite{billot2021synthseg,billot2020learning}.

\section{Conclusion}
Our research takes a significant step forward by synthesizing high-resolution 3D CT lung scans guided by detailed radiological and anatomical information. While GANs and cDPMs have set benchmarks, they come with inherent limitations, particularly when generating intricate chest CT scan details. Our proposed \ourmethod model addresses these challenges using a comprehensive dataset and a hierarchical training approach. Innovative architectural designs not only overcome previous constraints but also pioneer anatomy-conscious volumetric generation. 
Future work can leverage our model to enhance clinical applications.

\bibliography{tmi.bbl}
\bibliographystyle{IEEEtran}

\end{document}